\documentclass[runningheads]{llncs}
\usepackage[T1]{fontenc}
\usepackage{graphicx,verbatim,hyperref}
\usepackage{booktabs}
\usepackage{color}
\usepackage{amsmath}
\usepackage{subfig}

\usepackage{booktabs}
\usepackage{multirow}

\begin{document}
\title{Spline refinement with differentiable rendering}
%

\author{Frans Zdyb\inst{1}\orcidID{0009-0005-6062-7790} \and
Albert Alonso\inst{2}\orcidID{0000-0002-0441-0395} \and
Julius B. Kirkegaard\inst{1,2}\orcidID{0000-0003-0799-3829}}

\authorrunning{Zdyb et al.}

\institute{Deparment of Computer Science, University of Copenhagen, 2100, Denmark \and
Niels Bohr Institute, University of Copenhagen, 2100, Denmark
}



\maketitle

\begin{abstract}
Detecting slender, overlapping structures remains a challenge in computational microscopy.
While recent coordinate-based approaches improve detection, they often produce less accurate splines than pixel-based methods.
We introduce a training-free differentiable rendering approach to spline refinement, achieving both high reliability and sub-pixel accuracy.
Our method improves spline quality, enhances robustness to distribution shifts, and shrinks the gap between synthetic and real-world data.
Being fully unsupervised, the method is a drop-in replacement for the popular active contour model for spline refinement.
Evaluated on \textit{C. elegans} nematodes, a popular model organism for drug discovery and biomedical research, we demonstrate that our approach combines the strengths of both coordinate- and pixel-based methods.

\end{abstract}

\section{Introduction}

Automated detection and tracking methods form the basis for studying physiological behavior, drug effects, and motility in model organisms, enabling quantification of assays in biomedical research~\cite{hoffman2024artificial}.
Large animal models, such as mice and rats, are widely used for preclinical drug testing, allowing the quantification of complex physiological responses~\cite{alexandrov2015high,aljovic2022deep}.
Zebrafish provide a versatile system for high-throughput screening, as their transparent larvae enable real-time observation of drug-induced behavioral changes~\cite{rihel2012behavioral,miyawaki2020application}.
Fruit flies with their rapid generation time, serve as powerful genetic models for studying neural function and the effects of drugs on behavior~\cite{wang2020screening,qu2022easyflytracker}.
The detection and tracking of the behavior of these higher-order organisms are typically facilitated by landmark-based deep learning methods~\cite{perez2014idtracker,lauer2022multi,pereira2022sleap}, which enable precise monitoring of limb movements, wing beats, and overall locomotion patterns.
At the other scale, small model organisms such as the nematode \textit{Caenorhabditis elegans} offer a simple and cost-effective model for studying neuroactive compounds and locomotion-based toxicity~\cite{perni2017natural,ikenaka2019behavior,sohrabi2021high,kong2024behavioral}.
Due to their visually homogeneous appearance, landmark-based deep learning methods~\cite{lauer2022multi,pereira2022sleap} are not ideal for high-throughput analysis of \textit{C. elegans} assays.
Instead, pixel-based segmentation methods can often estimate nematode poses extremely accurately \cite{swierczek2011high,javer2018open,perni2018massively,cutler2022omnipose}.
However, these methods struggle when the images are at low resolution and worms appear thin, or when worms are occluded either due to self-overlap or due to imaging at high densities.
These challenges are overcome by spline-predicting methods \cite{hebert2021wormpose,alonsoFastDetectionSlender2023,weheliye2024improved}, which can track worms during self-coiling and at extreme densities.
These methods operate under a trade-off, however:
Their accuracy is not pixel-level-accurate and are highly adversely affected when applied for out-of-distribution inference.

Here, we propose an unsupervised model that uses differentiable rendering \cite{lake2015human} to refine spline predictions for pixel-level-accuracy.
Our method optimizes a reconstructed microscopy image, adjusting the background, textures, colors, spline shapes, and widths via differentiable rendering.
We compare this to the active contour model \cite{kassSnakesActiveContour1988}, an unsupervised refinement approach, which in contrast relies solely on feedback from the spline centerline, and therefore requires careful image preprocessing to work.
This model works per-spline and therefore does not generalize well to overlapping splines.
In contrast, in our approach, all splines are fitted simultaneously to an image.

We note that supervised deep learning refinement methods \cite{laube2018deep,wang2024b} are always faster and have greater detection accuracy.
However, these rely on pixel-perfect labels, which often are unavailable.
Such labels can be generated synthetically \cite{alonsoFastDetectionSlender2023}, where ground truth is fully controlled.
In that case, the challenge is that synthetic data introduces a gap between simulation and reality, as model assumptions and textures differ from real-world conditions.
Thus supervised refinement is possible for synthetic data, but bridging the reality-gap introduces the need for unsupervised refinement methods, such as the active contour model or the approach taken here.

\section{Method}
\begin{figure}[htb]
    \centering
    \includegraphics[width=\linewidth]{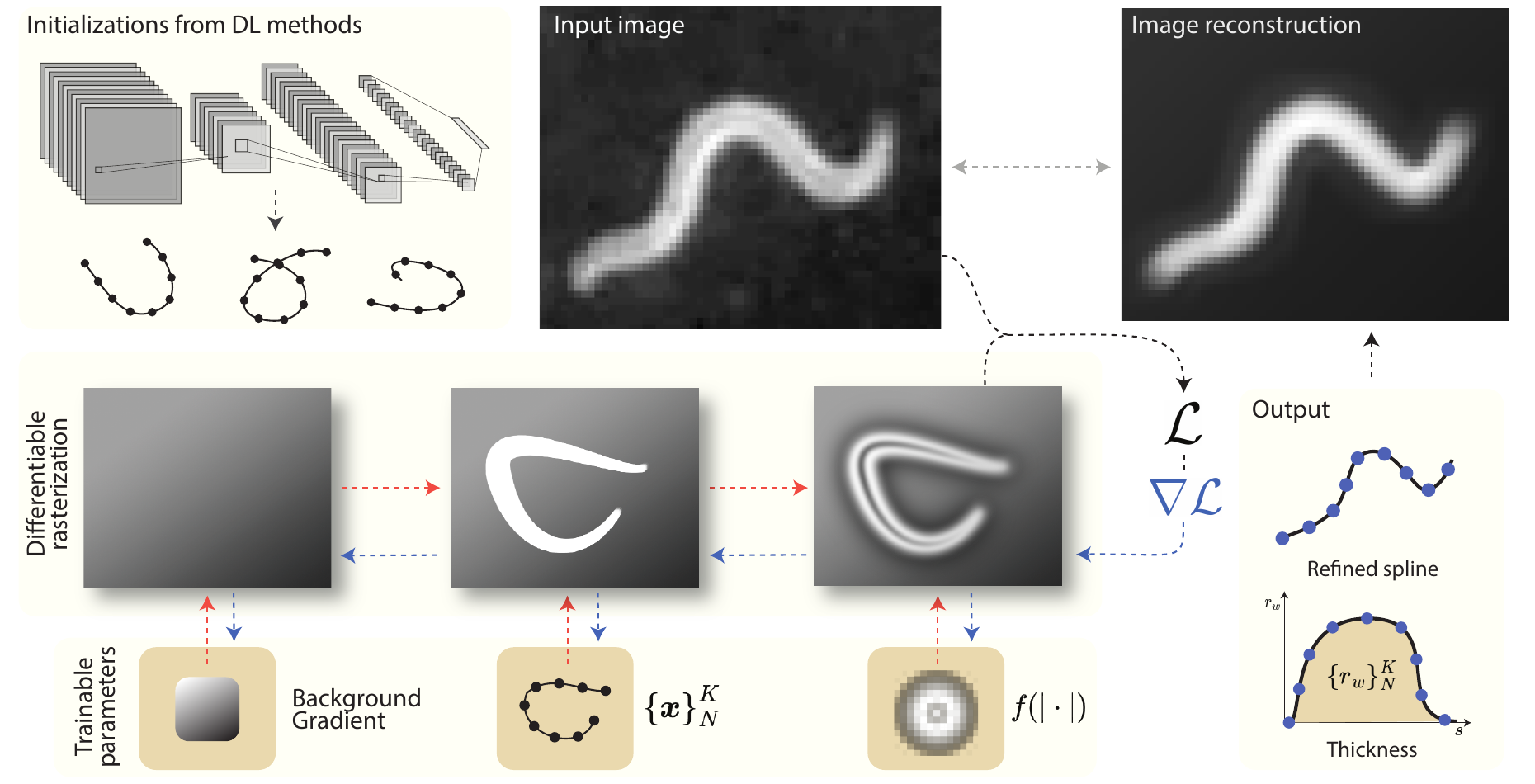}
    \caption{Overview of the spline refinement method using differentiable rendering.
    The process starts with initial spline guesses, potentially from deep learning models.
    Through differentiable rasterization (orange arrows), an image is rendered from spline control points and, background and texture adjustments.
    The loss $\mathcal{L}$ is computed by comparing the rendered and input images, and gradients (blue arrows) update trainable parameters (bottom).
    At convergence, the refined spline (bottom right) accurately reconstructs the input shape (top right), as well as provides auxiliary data e.g. spline thickness.
    }
    \label{fig:overview}
\end{figure}

The input to our model is one or more (potentially overlapping) splines that need to be refined.
These could, for instance, be the output of a deep learning detection model.
Our refinement approach, illustrated in Fig.~\ref{fig:overview}, reconstructs the input image through differentiable rendering by optimizing parameters to minimize the difference between the input and reconstructed images.
Our primary concern are the parameters that define the splines, but the optimization is performed jointly with the background colors and texture of the image.
Further, as our approach is fully differentiable, it is simple to regularize, ensuring e.g. smoothly curved spline (curvature regularization) or minor changes to total spline length.

We model the spline placement, spline fitting and rendering, along with background and texture generation, as an end-to-end differentiable program.
A set of parameters define the location, shape, and width of each spline. Specifically, for each set of $K$ knots $\{ \mathbf{x} \}^K = \{(x_1, y_1, w_1), ..., (x_K, y_K, w_K)\}$ we fit three natural cubic splines describing parametric curves of varying width,
\begin{equation}
\boldsymbol{r}^n : [0, 1] \rightarrow [0, \textsc{res}]^2 \times [0, 1] = (r_x(s), r_y(s), r_w(s)),
\end{equation}
where $(r_x, r_y)$ are coordinates in the image.
$r_w$ scales a global width parameter $W$ which is shared for all splines.
Additional parameters define background properties such as brightness, contrast and color gradient, and the texture of the image (Fig.~\ref{fig:overview}).
We have developed the method towards identifying splines whose texture is approximately constant along the spline, and we therefore learn only one (shared) blob texture.

\subsection{Differentiable rendering}
To render the reconstruction image, we sample a large set of interpolating $(x, y, w)$ points uniformly along each spline.
Each $(x, y)$ point is rendered onto a separate image by differentiable resampling of a learned blob image, which is scaled by $W \cdot w$.
The radially-symmetric blob image is produced by evaluating $f(z)$ in a small grid $\{(x_{ij}, y_{ij})\}$, where $z = \exp(-\sqrt{x^2 + y^2})$ and $f$ is a learned 1D spline function.
Summation over these scaled single-blob images produces a rendered spline.
The images generated for each spline are combined using summation and max operations, with a parameter controlling the interpolation between the two.
This allows for matching how overlapping splines interact visually.
A background is generated and added, and the result is convolved with a learned $3 \times 3$ convolutional filter, which allows for brightness/contrast matching and minor blurring.

\subsection{Optimization}
Our loss function is the reconstruction error between the input image $\mathbf{Y}$ and the generated image $\hat{\mathbf{Y}}$, 
\begin{equation}
    \mathcal{L} = ||\mathbf{Y} - \hat{\mathbf{Y}}||^2. \label{eq:reconloss}
\end{equation}
Akin to the active contour model \cite{kassSnakesActiveContour1988}, we regularize  the optimization with terms that exploit prior information (barred parameters):
\begin{align}
    \mathcal{L}_\text{reg} = \  \lambda_1  \sum_i^N (\ell_i - \bar{\ell}_i)^2 
   + \lambda_2 \sum_i^N C_i  + \lambda_3 \sum_n (\min_k w_k - \bar{w})^2  + \lambda_4 (W - \bar{W})^2.
\end{align}
Respectively, these terms regularize the spline length $\ell_i$ using the initial length $\bar{\ell}_i$ from the input knots, the curvature $C_i$ (smoothness) of the spline, and the minimum ($\min w_k W$) and maximum ($W$) point width.


\begin{figure}[tbh]
    \centering
    \includegraphics[width=\linewidth]{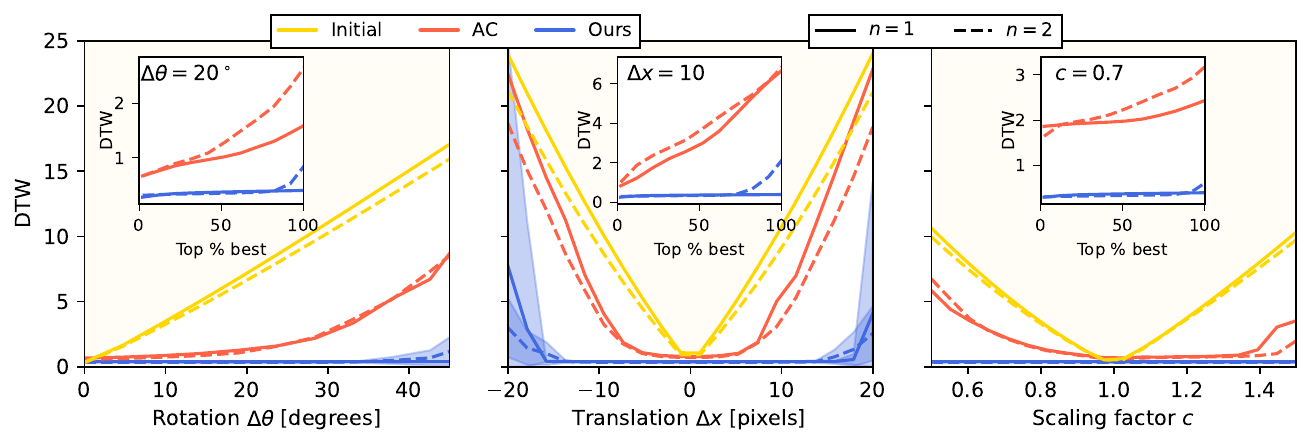}
    \includegraphics[width=\linewidth]{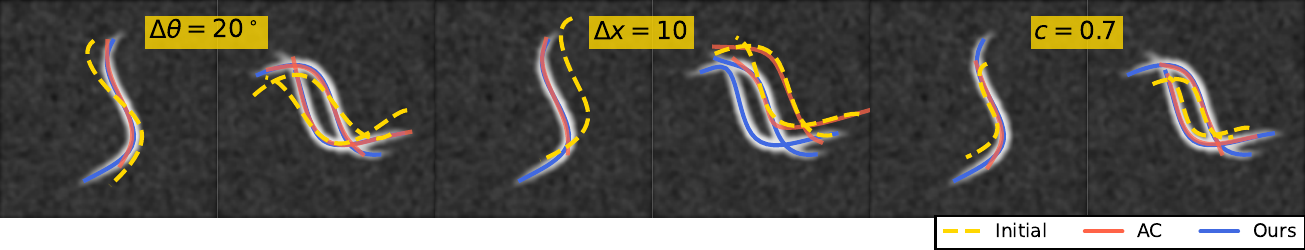}
    \caption{
    Robustness analysis of the proposed method on SOSB~(64 frames).  
    Average DTW after transformations: rotation (left), translation (middle), and scaling (right).
    The initial guess (yellow) is refined using active contour (red) or our method (blue).  
    Shaded areas indicate worsened performance. Main plots show the top 50\% performance, with insets displaying percentile variations.
    Some reconstruction examples are shown below.
    }
    \label{fig:quantitative}
\end{figure}

As our loss function is extremely non-convex, optimization is performed in three phases.
First, we freeze the spline parameters, and optimize the parameters of the background using Adam(1e-2).
Next, we optimize all parameters except the convolutional filter and the blob parameters, using Adam(1e-2) with cosine annealing.
Here we add Gaussian noise to the gradients, with a decaying variance, sampled with different random seeds, and select the parameters that achieve the lowest reconstruction loss.
Lastly, we finetune all parameters with Adam(1e-4).
While the choice of hyperparameters can be varied, the ordering of the optimization is crucial.
For instance, spline locations must be optimized prior to the blob image, as otherwise the blob risks converging towards the background color.
    


\section{Results}
\subsection{Evaluation Metrics}  
Centerline accuracy is often measured using mean squared error (MSE)~\cite{weheliye2024improved}, but it penalizes minor shifts that do not reflect significant spline differences.
Pixel-based metrics, such as the Jaccard Index~\cite{tahaMetricsEvaluating3D2015}, quantify overlap but are sensitive to image discretization, making them unsuitable for subpixel structures~\cite{cutlerOmniposeHighprecisionMorphologyindependent2022,alonsoFastDetectionSlender2023}. 
To avoid such issues, we here use average Dynamic Time Warping (DTW)~\cite{mullerDynamicTimeWarping2007}.
Before evaluation, both labels and refined splines are resampled to $N{=}100$ points using cubic interpolation.

\subsection{\textit{In-silico} Robustness Evaluation}
\subsubsection{Overlapping bodies dataset}
Access to high-quality labels for overlapping slender bodies is limited, obstructing quantitative evaluation~\cite{hebert2021wormpose,alonsoFastDetectionSlender2023,weheliye2024improved}.
Similarly, the majority of existing centerline prediction methods focus on single structures and struggle with overlaps~\cite{hebert2021wormpose,cutler2022omnipose}
Particularly for refinement studies, such as the present, labels need to be perfect to accurately reflect the improvements offered by refinement.
For that reason, we introduce a dataset of synthetic overlapping slender bodies~(SOSB)~\cite{overlappingDatasetSynthetic}, resembling microscopy scenarios with overlapping slender bodies~\cite{alonsoFastDetectionSlender2023,perni2018massively}.
This enables controlled comparisons and systematic evaluation as it provides perfect labels.

\subsubsection{Experiments}
Classic spline refinement methods, such as the active contour model~\cite{kassSnakesActiveContour1988}, adjust centerline positions using pixel gradients. They struggle when the gradients are weak or ambiguous, especially in microscopy images with occlusions and noise~\cite{alonsoFastDetectionSlender2023}.
Despite deep learning advancements, these methods remain popular due to their generality and ease of use, a trait shared by our proposed unsupervised differentiable approach.
We benchmark against active contour (AC) and assess robustness under perturbations, here exemplified by affine transformations of the centerline label: translation, rotation, and scaling~(Fig.~\ref{fig:quantitative}).
Similarly, we showcase the model's ability to refine PCA-compressed labels, which represent the prediction space of some deep learning models \cite{alonsoFastDetectionSlender2023,weheliye2024improved}.
The results shown in Fig.~\ref{fig:quantitative} and in Table~\ref{tab:overlap-quantitative} demonstrate improvement over classical refinement techniques.
For non-overlapping examples ($n = 1$), the method robustly recovers the labels even for extreme perturbations.
For overlapping splines ($n \geq 2$), failure examples occur at extreme perturbation levels, but the method remains robust even far outside the range of perturbations expected for spline initializations.

\begin{table}[tb]
    \centering
    \begin{tabular}{c|ccc|ccc|ccc|ccc}
    \toprule
    Number of & \multicolumn{3}{c|}{Rotation} & \multicolumn{3}{c|}{Displacement} & \multicolumn{3}{c|}{Scaling} & \multicolumn{3}{c}{PCA} \\
    Bodies $n$& \multicolumn{3}{c|}{$\Delta \theta=20^\circ$} & \multicolumn{3}{c|}{$\Delta x=10$} & \multicolumn{3}{c|}{$c=0.7$} & \multicolumn{3}{c}{$k=2$} \\
    \midrule
              & Initial & Ours & AC & Initial & Ours & AC & Initial & Ours & AC & Initial & Ours & AC \\
    1 & 7.14  & \textbf{0.48} & 0.85  & 9.38  & \textbf{0.49} & 2.27  & 5.33  & \textbf{0.53} & 2.39  & 5.88  & \textbf{0.52} & 1.44 \\
    2 & 7.14  & \textbf{0.43} & 1.51  & 9.33  & \textbf{0.40} & 4.37  & 5.31  & \textbf{0.45} & 2.65  & 5.71  & \textbf{0.40} & 1.79 \\
    3 & 7.18  & \textbf{1.63} & 2.49  & 9.36  & \textbf{0.78} & 5.62  & 5.30  & \textbf{1.39} & 3.16  & 5.69  & \textbf{1.62} & 2.37 \\
    \bottomrule
    \end{tabular}
    \caption{
    Comparison of average DTW distance between original and transformed label (initial), and after refinement using either our method or active contour.
    Best-performing in \textbf{bold}.
    Values show the top 50\% performance out of 256 frames for each $n$.
    }
    \label{tab:overlap-quantitative}
\end{table}

\subsection{Correcting \textit{in-vivo} Distribution Shifts}
Recent coordinate-based methods such as \texttt{de(ep)tangle}~\cite{alonsoFastDetectionSlender2023} and its newer extension \texttt{DeepTangleCrawl} (\verb|DTC|)~\cite{weheliye2024improved} have demonstrated well-performant spline detection and tracking of \textit{C.~elegans} in challenging scenarios, such as during overlapping and occlusions, and in noisy environments.
Despite their capabilities, these methods struggle in out-of-distribution situations where e.g. nematodes exhibit complex morphologies.
For example, tightly coiled or heavily occluded worms are difficult to label and are often underrepresented in training sets.
Such failures affect the robustness of the methods and can discourage their application in quantitative studies~\cite{deserno2025unsupervised}.

\begin{figure}[tb]
    \centering
    \includegraphics[width=\linewidth]{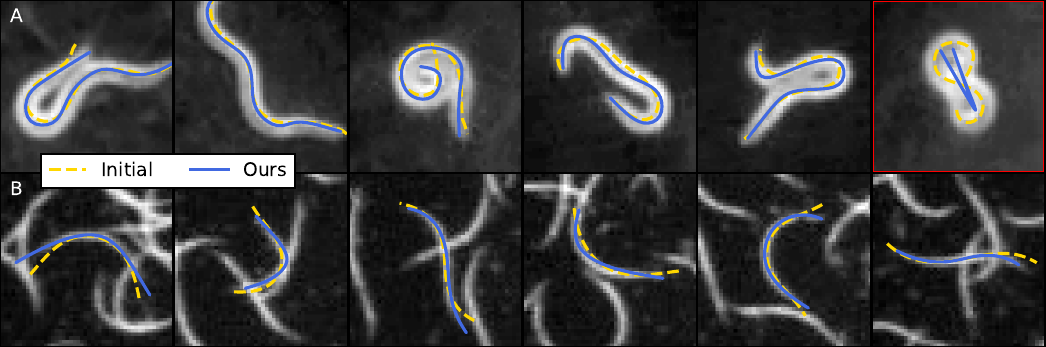}
    \caption{Examples of refinement (blue) of predictions (yellow) from (A)~\texttt{DeepTangleCrawl}~\cite{weheliye2024improved} and from (B)~\texttt{de(ep)tangle}~\cite{alonsoFastDetectionSlender2023} on low-density  \textit{C. elegans} 
 crawling experiments and high density swimming experiments, respectively. Example failure case is framed red.}
    \label{fig:qualitative}
\end{figure}

Here, we address such distribution shifts.  
For \verb|DTC|, distribution shifts arise because it relies on \verb|Tierspy|~\cite{javerOpensourcePlatformAnalyzing2018} to generate most of the labels, which performs well in most situations, but struggles when the boundaries of the nematodes are not well-defined.
Fig.~\ref{fig:qualitative}A demonstrates how our method corrects predictions for coiled worms, where initial predictions underperform, but refinement significantly improves accuracy.
In contrast, distribution shifts in \verb|de(ep)tangle|stem from training fully \textit{in-silico}.
This leaves a synthetic-to-real gap between the physics-based simulations and frame rendering and real worm recordings, which results in stiffer~(straighter) spline predictions than those observed~\cite{alonsoFastDetectionSlender2023}.
Fig.~\ref{fig:qualitative}B shows these examples and how our method can help resolve the issue even at this high density.
Note that while only one prediction is shown, we refine all visible splines simultaneously --- this is what allows refinements despite overlap.
These demonstrate the potential of our approach to not only refine predictions for coordinate-based deep learning models but also provide more accurate labels for future dataset curation.

\subsection{Assisted Differentiable Labeling}
\begin{figure}[tbh]
    \centering
    \includegraphics[width=\linewidth]{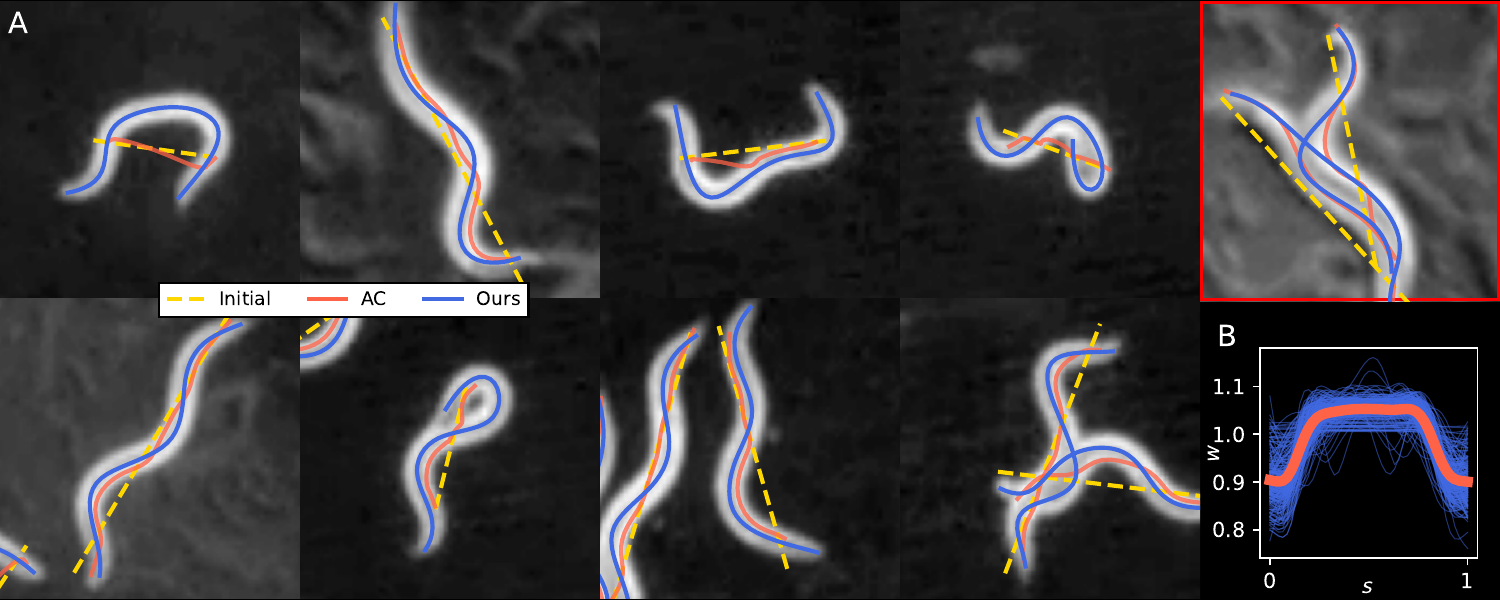}
    \caption{
    (A)~Refining low-information labels (straight lines).
    Example framed in red shows a failure case.
    (B)~Extracted scale parameter as a function of arc length, average in red.
    }
    \label{fig:applications}
\end{figure}

Labeling data for training supervised detection models is cumbersome and labor-intensive, particularly so for slender objects at high densities.
As our method can converge from extreme perturbations, it can enable an assisted labeling workflow in which quick first-pass labels, such as straight lines, are refined to full labels.
We show examples of this in Fig.~\ref{fig:applications}, and compare to \verb|DTC| labels in Table 2, demonstrating the potential of this idea.
While we mainly care about the splines, our method further gives auxiliary output.
For instance, the reconstruction loss can be used to assess whether a refinement was successful.
Likewise, our method outputs spline widths (Fig. ~\ref{fig:applications}B), which potentially could produce labels for pixel-based segmentation.
Similarly, being able to recreate the body of each spline independently can ease the process of edge detection even in situations where the bodies are obfuscated.

\begin{table}[tbh]
    \centering
    \begin{tabular}{c|ccc|ccc|ccc|ccc}
    \toprule
    Percentile & & 5 \% & & & 10 \% & & & 20 \% & & & 50 \% &\\
    \midrule
      &  Initial & Ours & AC & Initial & Ours & AC & Initial & Ours & AC & Initial & Ours & AC \\
    & 2.06  & \textbf{0.28} & 1.02  & 2.32  & \textbf{0.30} & 1.12  & 2.89  & \textbf{0.34} & 1.43  & 3.98  & \textbf{0.42} & 1.84 \\
    \bottomrule
    \end{tabular}
    \caption{
    Refining straight lines on \textsc{DTC} dataset, DTW percentiles. $n=1,\!000$.
    }
    \label{tab:straight-quantitative}
\end{table}

\section{Perspective}
In this paper, we have presented a method that refines centerline predictions to achieve pixel-level accuracy without requiring training.
Our approach relies on global reconstruction of the input image, leading to improved convergence compared to local centerline methods such as active contours.
More broadly, our results demonstrate the advantages of differentiable rendering-based matching, even when the reconstructed output only approximately resembles the input image.
Despite these benefits, our method has clear limitations.
For instance, failure cases highlight its sensitivity to the choice of regularization hyperparameters.
Concretely, for the used \textit{C. elegans} dataset, the method is generally successful in correcting both overlapping and coiling configurations, but may still fail due to regularization parameters being incorrectly tuned.
Parallel worms that appear as a single thick spline remain challenging.
Future work in the direction of 2D-splines with a temporal dimension could help resolve this.
Our approach is quite data-agnostic, but can be extended in a more data-specific direction, for instance by including prior knowledge relating to nematodes and the imaging process.
In particular, reformulating the differentiable program as a probabilistic program would allow for multi-modal priors, covering a much wider range of worm phenotypes without relying on tuning regularization hyperparameters. In addition, this could allow for quantifying uncertainty in the refinements.
Finally, refinement can enable self-supervised learning:
Any supervised model used to provide the initial spline can be fine-tuned on refined output, yielding better initial splines on future images.
Such a training loop, potentially guided by the image reconstruction loss, could be used to train deep supervised models with initially few but diverse training labels.

\subsubsection{Code availability.}
All code is available at \url{https://github.com/kirkegaardlab/splender}

\subsubsection{Acknowledgments.}
We thank Andr{\'e} EX Brown, Javier Rodriguez and Weheliye H Weheliye for providing us with the \textit{C. elegans} crawling data for the qualitative analysis.
This work has received funding from the Novo Nordisk Foundation, Grant Agreement No. NNF20OC0062047. 

\subsubsection{Disclosure of Interests.}
The authors have no competing interests to declare that are relevant to the content of this article.

\bibliographystyle{splncs04}

\end{document}